\documentclass[aps,pra,twocolumn,10pt]{revtex4-2}

\usepackage{xcolor}

\usepackage{epsfig,amsmath}
\usepackage{subfigure}
\usepackage{graphicx}
\usepackage{dcolumn}
\usepackage{stmaryrd}
\usepackage{mathrsfs}
\usepackage{pifont}
\usepackage{amsthm}
\usepackage{amssymb}
\usepackage{bm}
\usepackage{braket}
\usepackage{adjustbox}
\usepackage{latexsym}
\usepackage{color}
\usepackage{multirow}
\usepackage{xcolor}
\usepackage{babel}
\usepackage[ruled]{algorithm2e}
\usepackage{verbatim}
\usepackage{enumerate}
\usepackage{placeins} 
\usepackage{booktabs}

\usepackage{dsfont}
\usepackage{float} 
\usepackage{amsmath,amssymb}
\usepackage[utf8]{inputenc}
\usepackage{tabularx} 
\usepackage{url}           
\usepackage{hyperref}

\begin{document}

	\title{Realizing leakage elimination operator-based adiabatic speedup on a superconducting quantum processor}
	
	\author{
		Jia-Yi Fan\textsuperscript{1},\quad
		Kai-Yu Yuan\textsuperscript{2},\quad
		Zong-Yuan Ge\textsuperscript{1}, \quad
		Feng-Hua Ren\textsuperscript{3,$\#$}, \quad
		Zhao-Ming Wang\textsuperscript{1,4,5} 
	}
	\email{wangzhaoming@ouc.edu.cn}
	\thanks{$^\#$ renfenghua@qtu.edu.cn}
	\affiliation{$^1$College of Physics and Optoelectronic Engineering, Ocean University of China, Qingdao 266100, China}
	\affiliation{$^2$China Mobile (Suzhou) Software Technology Co., Ltd, Suzhou 215163, China}
	\affiliation{$^3$School of Sciences, Qingdao University of Technology, Qingdao 266033,  China}
	\affiliation{$^4$Engineering Research Center of Advanced Marine Physical Instruments and Equipment, Ministry of Education, Qingdao 266100,  China}
	\affiliation{$^5$Qingdao Key Laboratory of Advanced Optoelectronics, Qingdao 266100,  China}


	\begin{abstract}
		The slow evolution required for adiabaticity in adiabatic quantum computation renders the system vulnerable to environmental noise. Leakage elimination operator (LEO) control provides an effective strategy to realize adiabatic speedup over a short timescale, thus mitigating the noise impact. Despite extensive theoretical investigations, the realization of LEO-based adiabatic speedup on realistic superconducting quantum processors remains absent. In this work, we present such a realization on IBM superconducting quantum processors. We first characterize the trade-off between adiabaticity and noise accumulation by varying the total evolution time on both the Qiskit simulator and the \texttt{ibm\_marrakesh} processor, employing a comprehensive noise model that closely reproduces the experimental results. We then implement ideal LEO pulse derived for a closed system and achieve a significant enhancement of adiabatic fidelity within a short evolution time. To further improve the adiabatic fidelity, we refine the ideal LEO pulse via Bayesian optimization based on the comprehensive noise model. The optimized pulse yields a modest fidelity gain in simulation, yet on hardware it falls short of the ideal pulse under the present experimental conditions. Our work validates the feasibility of LEO‑based adiabatic speedup on a superconducting quantum processor and highlights the potential of LEO for noise‑aware adiabatic dynamics.
		
	\end{abstract}
	
	\maketitle
	
	\section{Introduction}
	Adiabatic evolution provides a powerful framework for adiabatic quantum computation~\cite{Nielsen2010}. By slowly varying the system Hamiltonian, the system remains approximately in its instantaneous ground state, thereby enabling robust state manipulation~\cite{Farhi2001,Albash2018}. However, environmental noise poses a major obstacle to realizing robust adiabatic processes~\cite{GarcaMolina2024}, and a trade-off emerges: adiabaticity requires infinitely long evolution times, yet prolonged exposure to the environment leads to accumulated decoherence, dissipation, and control errors, which progressively degrade the adiabatic fidelity~\cite{Gao2026}. This competition between adiabaticity and noise accumulation implies that, for given environmental parameters, an optimal total evolution time balances these two opposing effects~\cite{Wang2018,Smith2019}.
	
	As a promising strategy, adiabatic speedup has been developed to accelerate the adiabatic evolution, reaching the same final state on a much shorter timescale that is originally attainable only through a slow adiabatic passage~\cite{GueryOdelin2019}, thereby effectively suppressing the detrimental impacts of environment. Various approaches have been proposed, including counterdiabatic (transitionless) driving~\cite{Berry2009}, superadiabatic control~\cite{Giannelli2014}, Lewis–Riesenfeld invariant-based inverse engineering~\cite{Yu2025}, superadiabatic stimulated Raman adiabatic passage~\cite{Blekos2020}, and fast-forward scaling~\cite{Hatomura2024}, etc. Among these, the LEO method effectively suppresses quantum state leakage out of the target subspace by adding an LEO Hamiltonian to the system Hamiltonian~\cite{Wu2002,Jing2015,Wang2020a}. Using the Feshbach projection operator partitioning technique, analytical conditions for zero area periodic pulses have been derived in the high-frequency driving limit, for both sinusoidal and rectangular pulse shapes~\cite{Wang2020a}. Based on these conditions, LEO control has been theoretically shown to achieve almost exact state transfer in both closed~\cite{Wang2020a} and open quantum systems~\cite{Wang2020b}. The adiabatic speedups can also be realized by these ideal pulses~\cite{Wang2018} .
	
	Ideal pulses derived for closed quantum systems are generally not applicable in realistic noisy environments, especially when the system-bath couplings are strong~\cite{Wang2018,Wang2020b}. To address this, numerical optimization algorithms have been widely adopted to design robust control pulses that maximize fidelity under certain noise parameter, including the Adam algorithm~\cite{Xie2022,Xie2023,Yang2025}, gradient ascent pulse engineering (GRAPE)~\cite{Chen2025,Ding2019}, and the chopped random basis (CRAB) method~\cite{Muller2022}. Adam-based optimization has been used to realize adiabatic speedup in non-Markovian open qutrit systems~\cite{Xie2022} and high-fidelity quantum state transfer~\cite{Xie2023}. Notably, open-system GRAPE has been experimentally validated on superconducting quantum processors~\cite{Chen2025}. Recently, machine learning approaches, especially supervised learning with neural networks, have been employed to generate adaptive pulses for dynamic noise, further enhancing control robustness in adiabatic speedup~\cite{Gao2026,Zhang2025}. These advances create a clear path to implement theoretically proposed control protocols on real quantum hardware.
	
	IBM quantum processors have enabled fruitful demonstrations of quantum dynamics, including simulations of quantum many-body systems~\cite{Smith2019}, dynamical generation of decoherence free subspaces~\cite{Quiroz2024}, and molecular energy calculations~\cite{Aspuru2005}. Beyond these applications, various control and protection techniques have been experimentally validated on IBM hardware. For instance, dynamical decoupling has been shown to protect IBMQ qubits~\cite{sym15010062}, while counterdiabatic driving has been used to speed up adiabatic quantum computation~\cite{PhysRevApplied.15.024038}. Moreover, LEOs have been implemented for effective subspace protection~\cite{srep02075730}. Very recently, a comprehensive noise model accounting for Pauli errors, thermal relaxation, dephasing, and ZZ crosstalk has been developed to accurately reproduce the dynamics of high fidelity state transfer in spin chains~\cite{Ge2026}. 
	
	In this work, we realize LEO-based adiabatic speedup on a single-qubit system on IBM superconducting quantum processors. By varying the total evolution time on both the Qiskit simulator and the \texttt{ibm\_marrakesh} processor, we characterize the trade-off between adiabaticity and noise accumulation using a comprehensive noise model that matches experimental results well. Implementing ideal zero‑area rectangular LEO pulses, we observe a significant fidelity improvement from \(\sim 0.6\) to over \(0.95\) at the short evolution time \(T_{\mathrm{tot}} = \pi/2\). Based on this, we further refine the control pulse using Bayesian optimization~\cite{Fingar2024,Xie2022Bayesian} with the comprehensive noise model. Numerical simulations show a modest additional fidelity improvement over the ideal pulses, whereas on real quantum hardware, the results are comparable to those obtained with the ideal pulses or even slightly lower, due to device fluctuations and the mismatch between simulated and realistic noise models. Nevertheless, the simulation results still indicate the potential of pulse optimization for improving noise‑aware quantum control.

\section{Model and Methods}
	
	Consider a time-dependent Hamiltonian
	\begin{equation}
		H_0(t) = \frac{\Omega}{2} \left[ \cos(\omega t) \sigma_z + \sin(\omega t) \sigma_x \right].
	\end{equation}
	Here $\cos(\omega t)$ and $\sin(\omega t)$ can describe a field whose direction changes from $z$ to $x$ at constant angular velocity $\omega$, and $\Omega$ is the Rabi frequency. The instantaneous eigenvalues are $E_0 = -\Omega/2$, $E_1 = \Omega/2$. The eigenvectors are \cite{Wang201801}
	\begin{align}
		|E_0(t)\rangle &= -\sin\left(\frac{\omega t}{2}\right) |0\rangle + \cos\left(\frac{\omega t}{2}\right) |1\rangle, \\
		|E_1(t)\rangle &= \hphantom{-}\cos\left(\frac{\omega t}{2}\right) |0\rangle + \sin\left(\frac{\omega t}{2}\right)|1\rangle. 
	\end{align}
	
	For simplicity, suppose the system is initialized in the state $|E_0(t)\rangle=|1\rangle$, corresponding to the eigenstate of $\sigma_z$.
	According to the adiabatic theorem, the system remains in the instantaneous state throughout the evolution in the long time limit. At time $T_{\mathrm{tot}}=\pi/(2\omega)$, the state should be $(-\lvert 0 \rangle + \lvert 1 \rangle)/\sqrt{2}$.
	The adiabatic fidelity $F = \langle E_0(t)|\rho_s(t)|E_0(t)\rangle$ serves as a quantitative measure of adiabaticity, where $\rho_s(t)$ denotes the reduced density matrix of the system and $|E_0(t)\rangle$ is the instantaneous eigenstate of interest.

	Adiabatic speedup can be realized by introducing an LEO Hamiltonian into the system dynamics, which effectively suppresses the non-adiabatic transitions between different subspaces \cite{Wu2002}. The total Hamiltonian of the system under LEO control is given by
	\begin{equation}
		H(t) = H_0(t) + H_{\mathrm{LEO}}(t),
	\end{equation}
	where the LEO Hamiltonian takes the form
	\begin{equation}
		H_{\mathrm{LEO}}(t) = c(t) |E_0(t)\rangle \langle E_0(t)|.
	\end{equation}
	Here $c(t)$ represents the control function that can be implemented with a sequence of tailored control pulses \cite{Wang201801}. For closed quantum systems, the ideal zero-area pulse conditions can be derived using the P-Q partitioning technique \cite{Wang2020a}. A variety of pulse profiles have been investigated in previous studies, including rectangular and sinusoidal pulses \cite{Wang2020a}.
	For the case of rectangular pulses
	\begin{equation}
		c(t) = \begin{cases}
			I, & n\tau < t < (n+1)\tau,\quad n\ \text{even},\\
			-I, & \text{otherwise},
		\end{cases}
		\label{eq6}
	\end{equation}
	where $I$ is the pulse intensity and $\tau$ is the half period of the pulse. The pulse conditions are \cite{Wang2020a}
	\begin{equation}
		I \tau = 2\pi m \quad (m = 1,2,3,\ldots).
		\label{eq7}
	\end{equation}
	
	We construct quantum gate circuits to simulate the adiabatic evolution of a single qubit under three scenarios: noiseless evolution, noisy evolution without control, and noisy evolution with LEO control. All circuits are implemented on the Qiskit platform using the Suzuki–Trotter decomposition. The latter two scenarios are additionally validated on IBM quantum processors. Detailed circuit constructions and gate decompositions are provided in Appendix~\ref{app:circuit}.
	
	The adiabatic evolution is realized using a sequence of $R_X$ and $R_Z$ rotations, where the continuous time evolution is approximated via a multistep Trotter expansion.
	The LEO control Hamiltonian is incorporated by applying projective operations at each Trotter time step~\cite{Smith2019}.
	The full LEO-based circuit is then transpiled into the native gate set of the \texttt{ibm\_marrakesh} processor,
	$\{\text{id}, \text{cz}, \text{rx}, \text{rz}, \text{sx}, \text{x}, \text{rzz}\}$. This transpilation improves agreement with realistic hardware noise and ensures compatibility with the backend architecture.

	To accurately capture the realistic noise in quantum hardware, we utilize a comprehensive noise model for our simulations~\cite{Ge2026}. These noise channels are calibrated using the official noise parameters and gate duration data provided by IBM quantum processors, ensuring that the simulations of the dynamics closely match those observed on actual chip operation. Details of the noise model, including parameter settings and channel construction, are provided in Appendix~\ref{app:noise}.
	
	\section{Adiabatic Evolution without pulse control}
	\begin{figure}[htbp]
		\centering
		\subfigure[]{
			\includegraphics[width=0.97\linewidth]{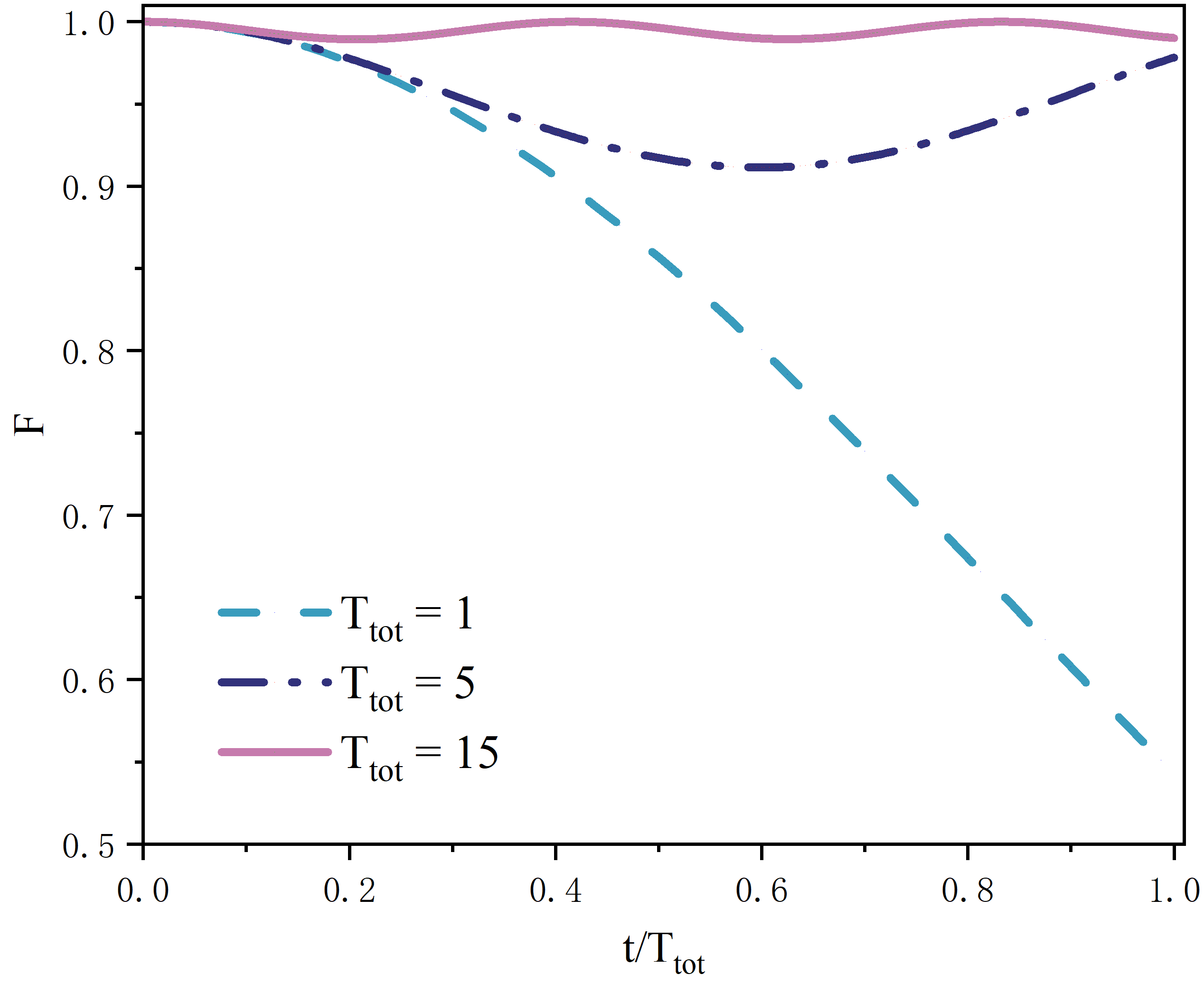}
			\label{fig:1a}
		}
		
		\subfigure[]{
			\includegraphics[width=\linewidth]{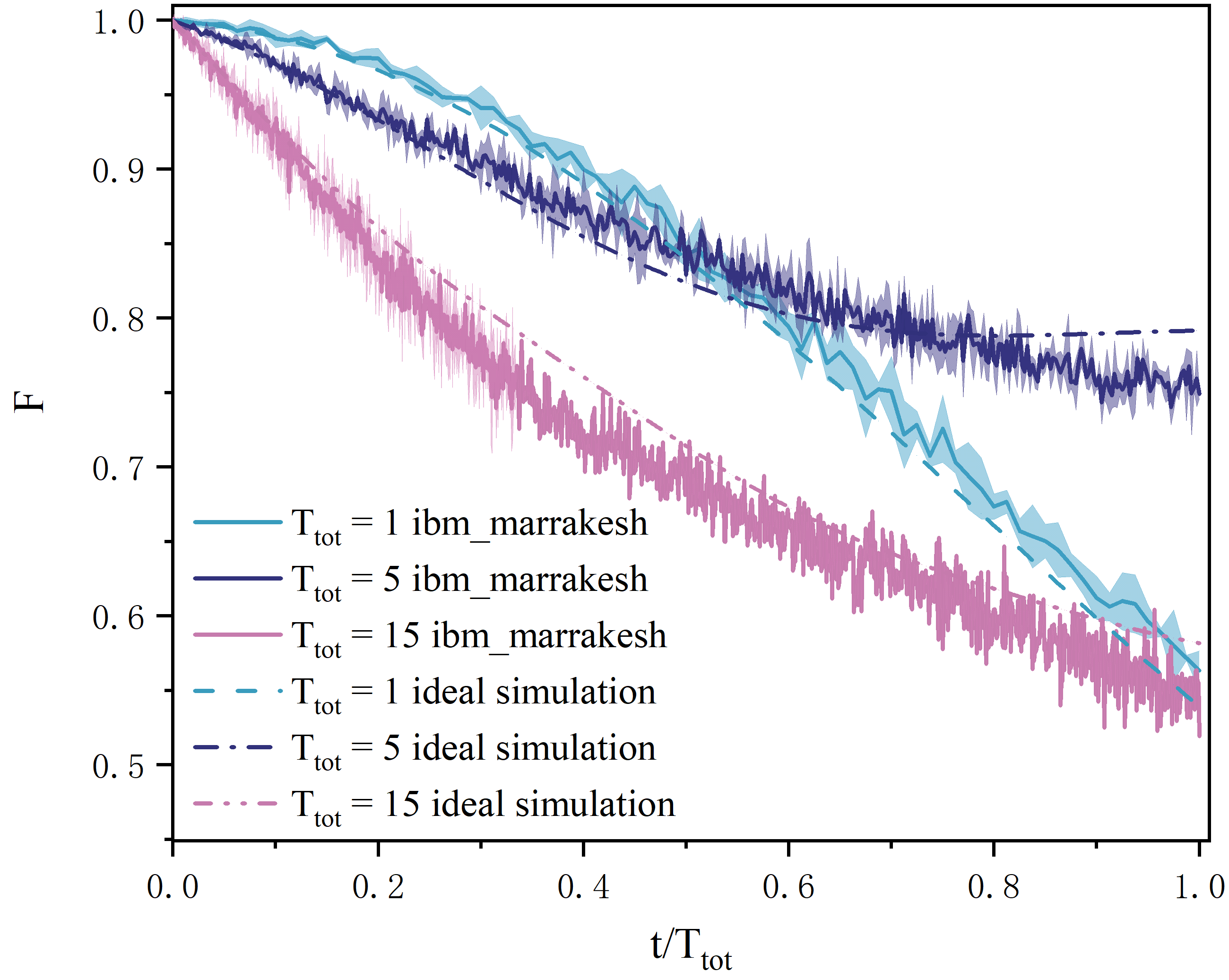}
			\label{fig:1b}
		}
		\caption{(a) Simulated fidelity versus rescaled time $t/T_{\mathrm{tot}}$ for different $T_{\mathrm{tot}}$ without noise. (b) Comparison of adiabatic fidelity between simulation with a comprehensive noise model (including \(T_1,T_2\) and depolarizing noise) and experimental data on the \texttt{ibm\_marrakesh} for \(T_{\mathrm{tot}} = 1, 5, 15\). For \texttt{ibm\_marrakesh}, the curves represent averages over three experimental runs, and the shaded regions indicate the error bars.}
		\label{fig:1}
	\end{figure}
	
	We first study the adiabatic evolution of a single-qubit from $\lvert 1 \rangle$ to $(-\lvert 0 \rangle + \lvert 1 \rangle)/\sqrt{2}$ without control. The Rabi frequency is set to $\Omega = 1$, and we take $\omega = \pi/(2T_{\mathrm{tot}})$. The number of Trotter steps per unit time is fixed at 80.

	Figure~\ref{fig:1} plots the adiabatic fidelity $F$ as a function of the rescaled time $t/T_{\mathrm{tot}}$ for different $T_{\mathrm{tot}}$. Results from both the \texttt{ibm\_marrakesh} quantum processor and Qiskit simulations (with identical parameters and a comprehensive noise model~\cite{Ge2026}) are presented, enabling a direct comparison. Noiseless simulations are shown in Fig.~1(a), while noisy simulations and the experimental data from \texttt{ibm\_marrakesh} are presented in Fig.~\ref{fig:1}(b). We first examine the simulation results. In Fig.~\ref{fig:1}(a), for a short $T_{\mathrm{tot}} = 1$, the dynamics lies in the non‑adiabatic regime and the fidelity drops to $0.563$. The same final fidelity is observed in the noisy simulation of Fig.~\ref{fig:1}(b) at $T_{\mathrm{tot}} = 1$, indicating that noise has little effect over such a short evolution. Increasing $T_{\mathrm{tot}}$ to 5 raises the noiseless fidelity to 0.978, whereas the noisy simulation gives $0.749$ and the noise begins to degrade the fidelity. When $T_{\mathrm{tot}}$ is further extended to 15, the noiseless case (Fig.~\ref{fig:1}(a)) reaches $F \approx 1$, firmly in the adiabatic regime.  However, in Fig.~\ref{fig:1}(b) the fidelity falls to $0.519$ due to the accumulation of noise, even lower than the value at $T_{\mathrm{tot}}=5$. This clearly demonstrates the trade-off we mentioned earlier: in the presence of environmental noise, simply prolonging the evolution time to improve adiabaticity has intrinsic limitations~\cite{Gao2026}. 
	
	Comparing the noisy simulation with the experiment from the \texttt{ibm\_marrakesh} in Fig.~\ref{fig:1}(b), we see that the overall trend is consistent. However, as $T_{\text{tot}}$ increases, the experimental data exhibit strong oscillations. This arises from the fact that a larger $T_{\text{tot}}$ leads to deeper circuits, which become too demanding to execute reliably, forcing a reduction in the number of shots and thereby exacerbating the oscillations. For the \texttt{ibm\_marrakesh} processor, each data point is averaged over three experimental runs, and the shaded regions represent the error bars. We also observe a deviation between the simulated and measured fidelity, likely because real quantum hardware is subject to additional noise sources, including gate errors, measurement errors, and control noise. In the simulations, a thermal relaxation noise channel is applied after each quantum gate, with characteristic parameters $T_1=180.86\,\mu\mathrm{s}$, $T_2=94.15\,\mu\mathrm{s}$, and a single-qubit gate duration of $36\,\mathrm{ns}$. The noise parameters may vary slightly in subsequent experiments. Additionally, a weak depolarizing channel is included as a phenomenological correction to capture residual errors not fully described by thermal relaxation and dephasing. Detailed information is provided in the Appendix~\ref{app:noise}.

\begin{figure}[htbp]
		\centering
		\includegraphics[width=1\columnwidth]{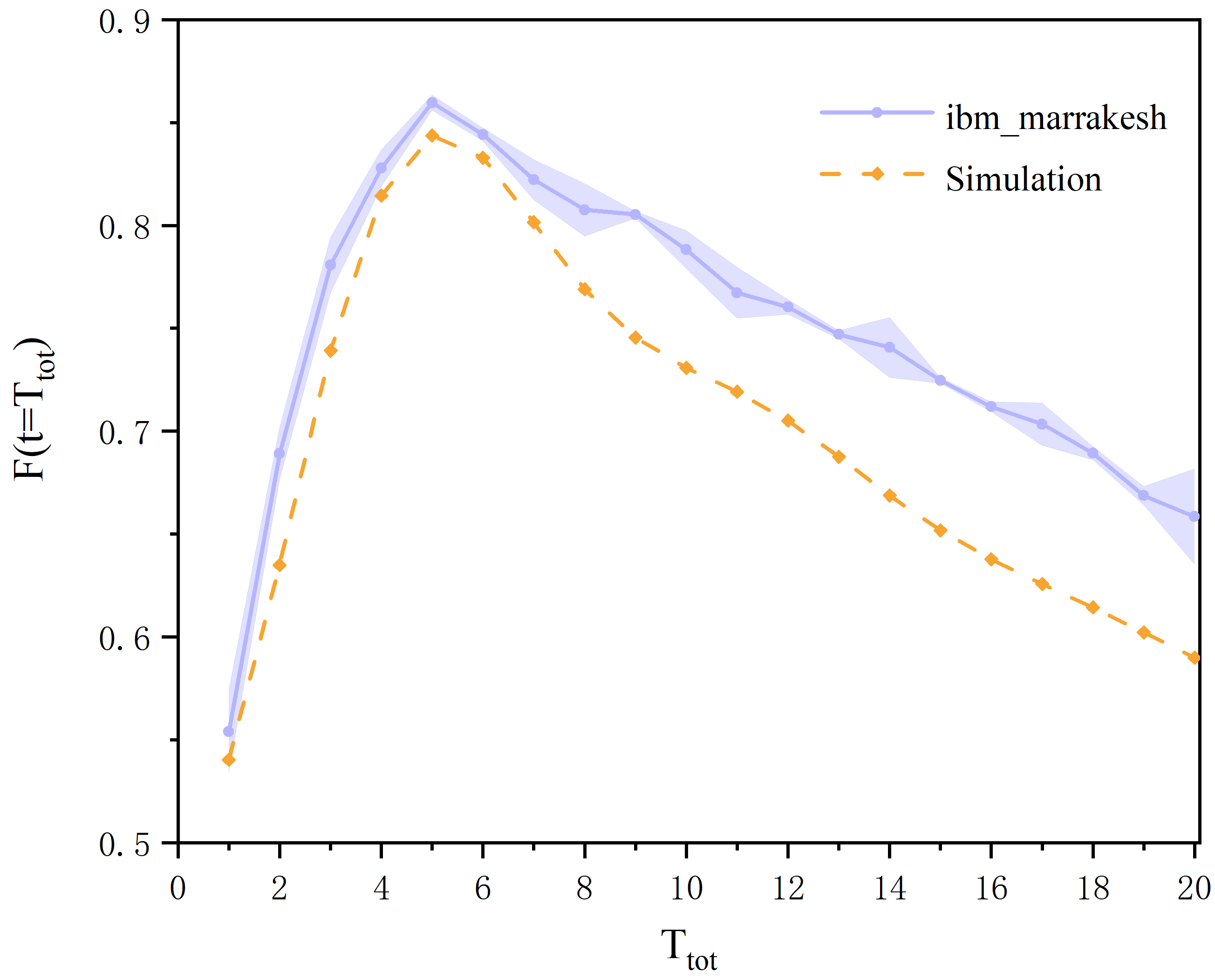}
		\caption{Final fidelity $F(t=T_{\mathrm{tot}})$ versus total evolution time $T_{\mathrm{tot}}$ from the \texttt{ibm\_marrakesh} processor (averages over three runs; shaded regions indicate error bars) and Qiskit simulations with a comprehensive noise model.}
		\label{fig:2}
	\end{figure}

	Due to the trade-off between the adiabaticity and short evolution time, an optimal $T_{\mathrm{tot}}$ exists~\cite{Wang2018adiabatic}. Consequently, the final fidelity depends on $T_{\mathrm{tot}}$, as shown in Fig.~\ref{fig:2}. Quantum circuits with different $T_{\mathrm{tot}}$ are executed on the \texttt{ibm\_marrakesh} and in simulations. The results indicate that as $T_{\mathrm{tot}}$ increases, the fidelity first rises and then falls, reaching a peak at $T_{\mathrm{tot}} = 5$. Figure~\ref{fig:2} clearly illustrates the underlying trade-off: for short total evolution times, adiabaticity is low, but environmental effects are weak. As $T_{\mathrm{tot}}$ increases, adiabaticity improves and fidelity grows, while environmental noise becomes increasingly destructive. Beyond the optimal $T_{\mathrm{tot}}$, environmental degradation dominates, and further extending the evolution time reduces the fidelity. 
	
	The simulation results agree well with the \texttt{ibm\_marrakesh} data for short $T_{\mathrm{tot}}$, but a slight deviation appears when $T_{\mathrm{tot}}$ exceeds the optimal value of $5$. The experimental fidelity is slightly higher than the simulated one. This discrepancy arises because the actual noise parameters on \texttt{ibm\_marrakesh} are complex and vary from qubit to qubit, making the parameters used in our noise model imprecise. Additionally, the circuit after mapping to the basis gates of \texttt{ibm\_marrakesh} differs somewhat from the circuit used in our simulations, which also contributes to the simulated fidelity being slightly lower than the experimental value.

	\section{Adiabatic Speedup under LEO Control}
	As discussed above, for short $T_{\mathrm{tot}}$, the influence of environmental noise is relatively weak. In this regime, LEO control can be employed to prevent the system from one subspace to other spaces~\cite{Wu2002} and effective adiabatic speedup can be realized~\cite{Wang2018adiabatic}. Here, we use the rectangular pulses in Eq.~(\ref{eq6}) together with the ideal pulse conditions in Eq.~(\ref{eq7}) to implement the adiabatic speedup.
	
	Within each Trotter time step, the quantum gates composing $H_0$ are first applied, followed by the LEO control gate that projects the system state onto the instantaneous target state at the corresponding time step.
	In the quantum circuit implementation, the LEO control gate is constructed from the projector-based evolution
	\begin{equation}
		U_{\mathrm{LEO}} = I + \bigl(e^{-i\phi} - 1\bigr) |E_0(t)\rangle \langle E_0(t)|,
	\end{equation}
	where $\phi = \int_{0}^{t} c(s)\,ds$ is the accumulated phase. For the rectangular pulse in Eq.~(\ref{eq6}), the phase
	\begin{equation}
		\phi(t) = \begin{cases}
			It, & n\tau < t < (n+1)\tau,\quad n\ \text{even},\\
			I(2\tau-t), & \text{otherwise}.
		\end{cases}
		\label{eqphase}
	\end{equation}
	
\begin{figure}[htbp]
		\centering
		\includegraphics[width=1\columnwidth]{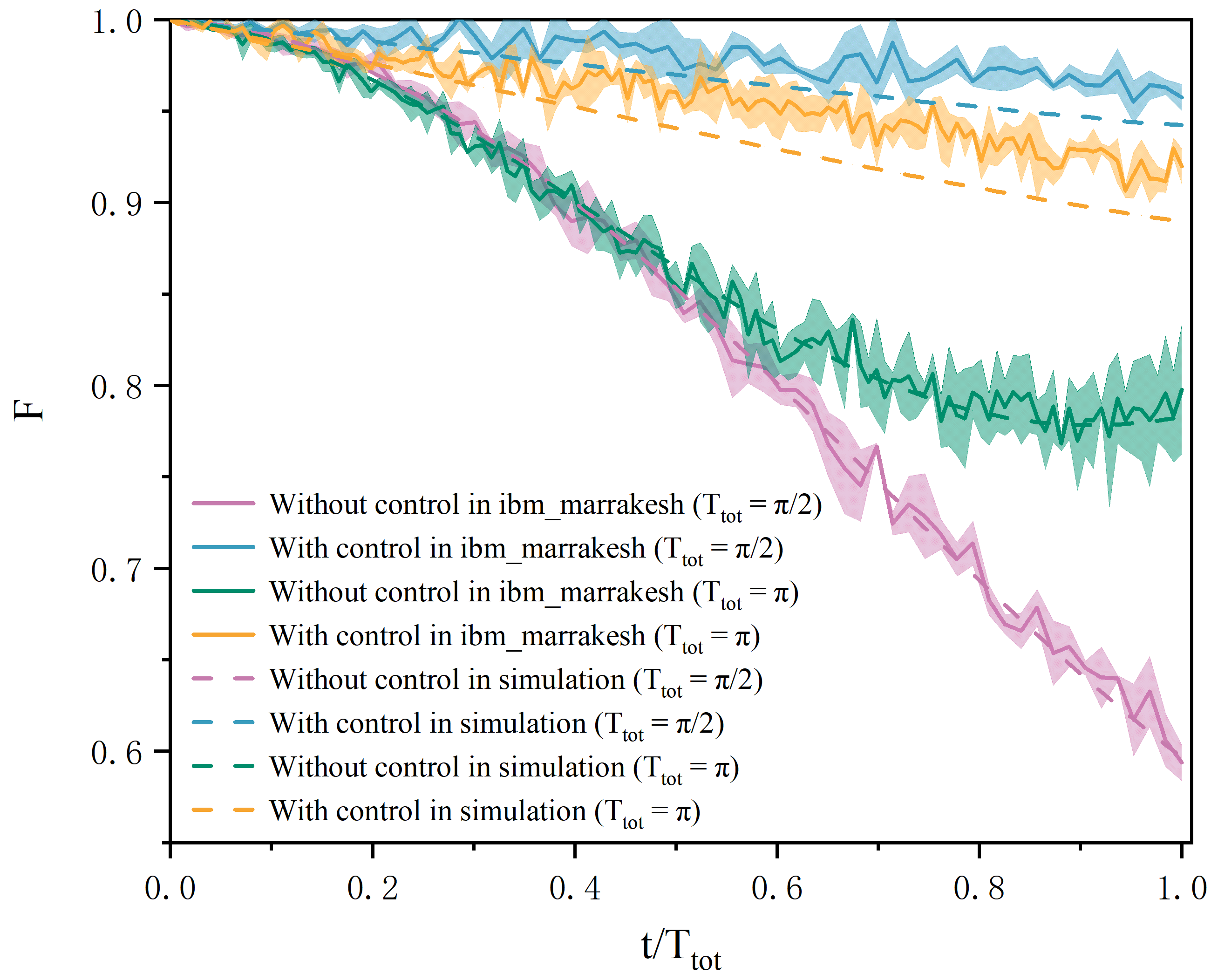}

\caption{Fidelity $F$ versus rescaled time $t/T_{\mathrm{tot}}$ for simulations with a comprehensive noise model and the \texttt{ibm\_marrakesh} processor, at $T_{\mathrm{tot}} = \pi/2$ and $\pi$, both with and without LEO control. For \texttt{ibm\_marrakesh}, curves are averages over three runs, shaded regions denote error bars.}
		\label{fig:3}
	\end{figure}

	After compilation, this control gate is decomposed into the native gate set of the quantum hardware and then inserted after each Trotter step to achieve the desired LEO control. Figure~\ref{fig:3} illustrates the performance of LEO control in both simulations and on \texttt{ibm\_marrakesh} processor, comparing the fidelities achieved with and without LEO control. Tests are performed for two total evolution times: $T_{\mathrm{tot}} = \pi/2$ and $T_{\mathrm{tot}} \approx \pi$, with $63$ and $126$ Trotter steps, respectively. For $T_{\mathrm{tot}} = \pi/2$, the evolution time is short, resulting in a low fidelity. However, introducing LEO control significantly improves the fidelity in both simulations and hardware experiments, raising it from approximately $0.6$ to over $0.96$, thereby confirming the feasibility of the proposed adiabatic speedup strategy. For $T_{\mathrm{tot}} = \pi$, the longer evolution time increases circuit depth, which amplifies noise accumulation and degrades the performance of LEO control. Nevertheless, LEO control still enhances the fidelity compared to the uncontrolled case.
	
	For both evolution times, the hardware results with LEO control exhibit higher fidelities than the numerical simulations, yet the overall trends remain in good agreement. Even with LEO control, the fidelity does not reach unity. Typical values are $F(t=\pi/2)\approx0.96$ and $F(t=\pi)\approx0.92$ due to unavoidable environmental noise. In this demonstration, the LEO pulse parameters are set to $I = 40$ and $\tau = \pi/20$, satisfying the condition in Eq.~(\ref{eq7}). The noise parameters used in the simulation are consistent with those calibrated from the quantum processor.

	\section{Improving fidelity via pulse optimization}
	As illustrated in Fig.~\ref{fig:3}, adiabatic fidelity can be improved via LEO pulse control. However, the used ideal pulses were originally derived for closed quantum systems. For weak system-bath couplings, the system can be considered nearly closed, and in this regime ideal pulse control works well. As the coupling strength increases, the control gradually loses its effectiveness~\cite{Wang2018adiabatic}.
	
	Numerically, conventional optimization algorithms have been adopted to synthesize noise-robust pulses that are resilient against time-independent noise~\cite{Yang2025}. Neural-network-based pulse design has also been explored to counteract time-dependent noise~\cite{Gao2026,Shi2024,Stern2021}. In this work, we try to perform pulse optimization to further boost adiabatic fidelity on the \texttt{ibm\_marrakesh} processor. The basic idea is as follows: first, optimal control pulses are constructed under specific environmental parameters using a comprehensive noise model. Then, the optimized pulses are implemented on \texttt{ibm\_marrakesh} to verify their performance.
	
	To efficiently parameterize the control pulse while naturally preserving its zero-area property, we directly optimize the control function $c(t)$ using a Fourier-series representation. Specifically, the total evolution time is divided into $M$ optimization segments. Within each segment, $c(t)$ is expanded as a truncated odd-harmonic Fourier series
	\begin{equation}
		c^{(p)}(t) = \sum_{j=0}^{K-1} a_j^{(p)} \sin\big[(2j+1)\Omega_\tau t\big],
	\end{equation}
	where $\Omega_\tau = 2\pi/\tau$, $K$ is the number of harmonics, and $\{a_j^{(p)}\}$ are the Fourier coefficients to be optimized within the $p$-th segment. The use of odd-harmonic sine components automatically suppresses the average pulse area over each oscillation period~\cite{Gao2026}. The optimized discrete accumulated phases $\phi$ are then obtained via Trotterized integration: $\phi_k \approx c_k \Delta t$. The Fourier coefficients are optimized using a greedy sequential segment-wise Bayesian optimization strategy ~\cite{frazier2018} (see Appendix~\ref{app:optimization} for details). This parameterization significantly reduces the dimensionality of the optimization problem while preserving the oscillatory zero-area structure of the original control pulses. Moreover, the Fourier representation provides a smooth and physically motivated waveform ansatz, making the optimization more robust against noise and numerical fluctuations. The segment-wise optimization strategy further improves computational efficiency by decomposing the high-dimensional global optimization problem into a sequence of low-dimensional local optimization tasks.
	
	\begin{figure}[htbp]
		\centering
		\subfigure[]{
			\includegraphics[width=\linewidth]{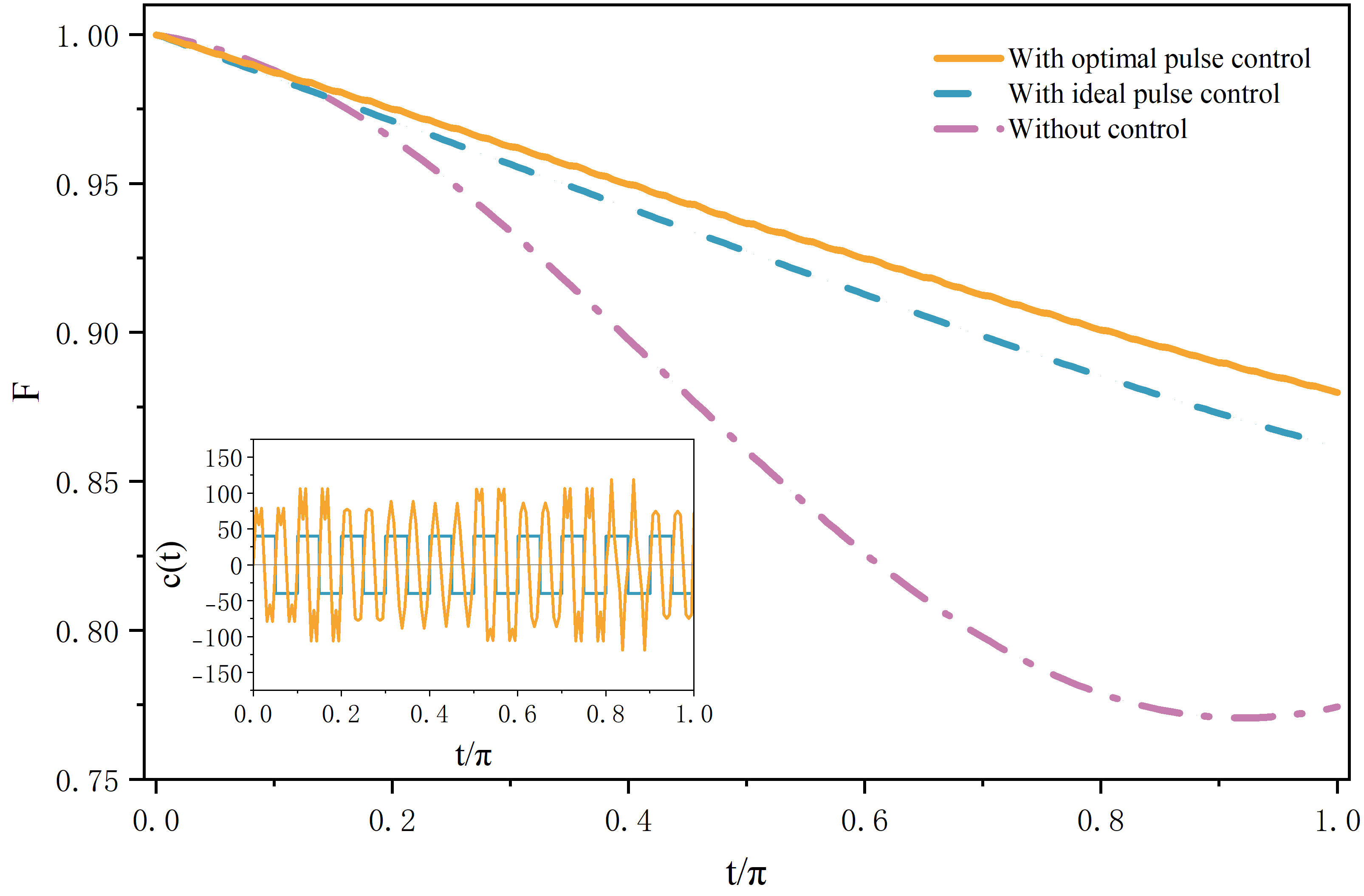}
			\label{fig:4a}
		}
		
		\subfigure[]{
			\includegraphics[width=\linewidth]{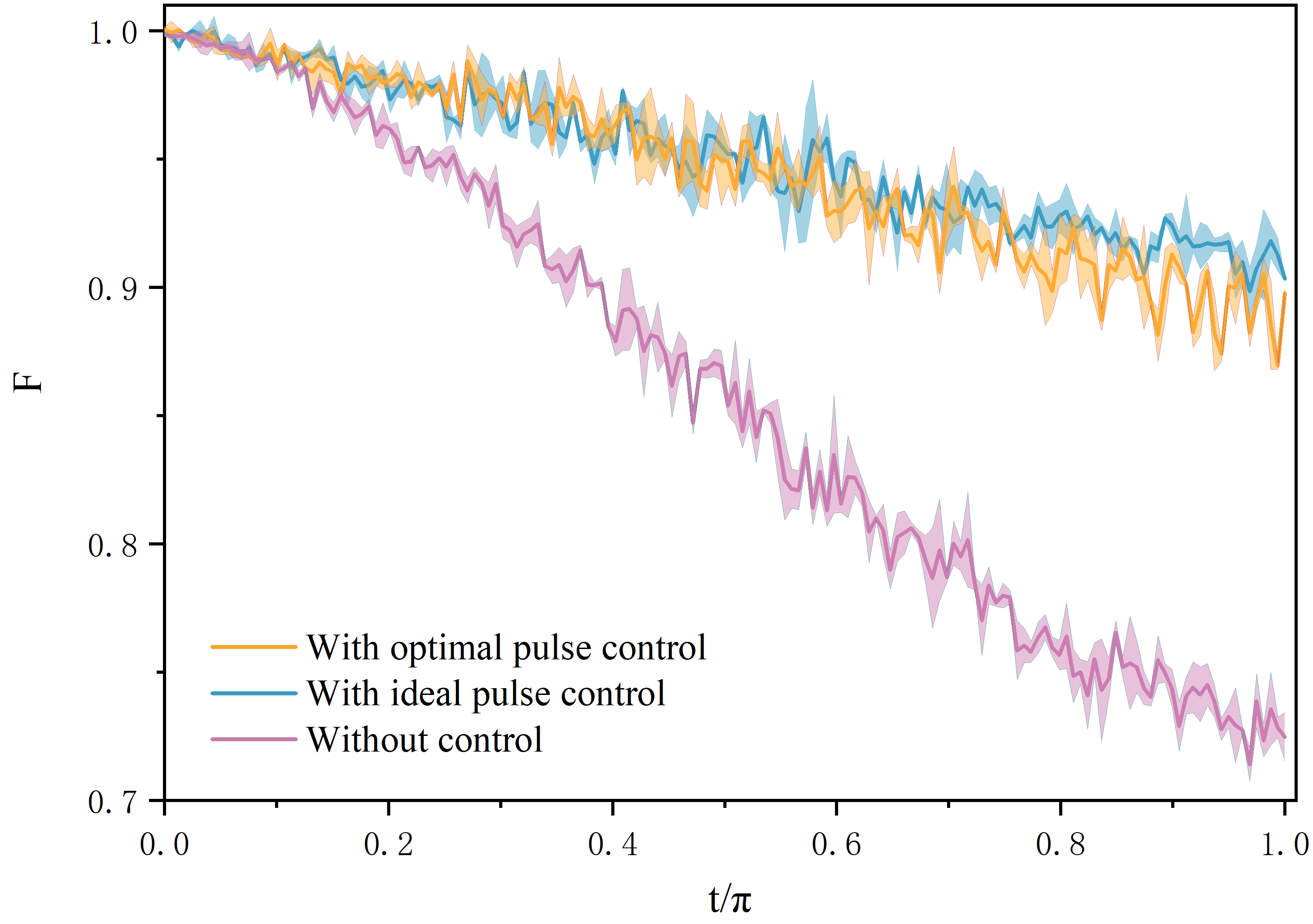}
			\label{fig:4b}
		}
		
		\caption{(a) Simulated fidelity $F$ versus $t/\pi$ without control, with ideal pulse control, and optimized pulse control. Inset shows the corresponding ideal and optimized LEO pulse sequences. (b) Fidelity $F$ versus $t/\pi$ on the \texttt{ibm\_marrakesh} processor with ideal and optimized LEO pulse control. Curves are averages over three experimental runs; shaded regions are error bars. $T_{\mathrm{tot}}=\pi$. For the ideal pulse, $I=40$, $\tau=\pi/20$.}
		\label{fig:4}
	\end{figure}

	Figure~\ref{fig:4a} presents Qiskit simulation results of the fidelity $F$ versus time $t/\pi$ under three cases: no control, ideal control, and optimized control. $T_{\mathrm{tot}}=\pi$. From Fig.~\ref{fig:4a}, the fidelity drops to about $0.77$ in the free evolution, while under an ideal pulse with $I=40$ and $\tau=\pi/20$, it is enhanced to $F=0.86$. Optimizing the ideal pulse further improves the simulated fidelity by approximately $1.83\%$. The corresponding waveforms (ideal and optimized) are shown in the inset of Figure~\ref{fig:4b}.
	
	We then implement the optimized pulse sequence on \texttt{ibm\_marrakesh}, and the experimental results are plotted in Figure~\ref{fig:4b}. Although the optimized pulse exhibits a modest additional improvement in numerical simulations, the hardware fidelity remains very close to that obtained with the ideal pulse and is slightly lower in the present experiment. This result likely indicates that the optimized waveform does not fully match the realistic hardware noise. In particular, real superconducting processors involve additional error sources such as readout errors~\cite{Hicks2022}, crosstalk~\cite{Winick2021}, leakage~\cite{Werninghaus2021}, coherent control errors~\cite{Berberich2024}, and pulse-level imperfections~\cite{Wiening2025} that are difficult to accurately reproduce in simulations. Nevertheless, the additional improvement observed in the simulations suggests that waveform optimization still has the potential to further enhance adiabatic control performance when more accurate hardware-aware noise modeling and pulse optimization strategies are employed. Moreover, the original ideal zero-area pulse already provides strong adiabatic protection under the present experimental conditions, leaving only limited room for further improvement through waveform optimization. Finally, the additional improvement predicted by simulations is relatively small and can be partially masked by hardware fluctuations including shot noise and calibration drift~\cite{Xu2025,Gaye2025}. Note that the estimated fidelity may occasionally exceed unity. This is primarily due to statistical fluctuations arising from finite-shot measurements of $\langle X\rangle$ and $\langle Z\rangle$, combined with the error-mitigation procedures employed by the IBM Runtime Estimator. As the reported values are mitigated expectation-value estimates rather than directly measured probabilities, they are not strictly bounded by $1$. Such deviations are small and do not affect the observed fidelity trends or the conclusions drawn in this work.
	
	\section{CONCLUSIONS}
	
	In this work, we have demonstrated adiabatic speedup using LEOs on both an IBM quantum processor (\texttt{ibm\_marrakesh}) and in Qiskit simulations. Using a comprehensive noise model that includes $T_1$ relaxation, $T_2$ dephasing, and gate errors, our simulations show excellent agreement with experimental results on superconducting hardware. For single‑qubit adiabatic evolution, an optimal total time $T_{\mathrm{tot}}$ arises from the trade‑off between the slow evolution required by the adiabatic theorem and the fast evolution needed to mitigate noise. Beyond this optimum, fidelity decreases monotonically with increasing $T_{\mathrm{tot}}$, indicating that noise disrupts adiabaticity.
	
	We then introduce ideal LEO‑based pulses derived for the closed‑system to accelerate the adiabatic process. Both simulations and experiments confirm the effectiveness of this approach. To further enhance performance, we design optimized control pulses tailored to the noisy environment using the same noise model, starting from the ideal pulse. Numerical simulations show a modest additional fidelity improvement under these noise‑aware pulses, but hardware results remain comparable to those obtained with the ideal pulses or even slightly lower in the present experiment. Our results confirm the effectiveness of zero‑area pulse control, which holds promising potential for high‑quality quantum information processing. The present work is limited to a single‑qubit system and multi‑qubit system should be studied in a realistic environment in the future. 
	\section*{acknowledgements}
	This work is supported by the  Natural Science Foundation of Shandong Province (Grant No. ZR2024MA046).
	
	\section*{DATA AVAILABILITY}
	The data are available from the corresponding author upon reasonable request.
	
	\appendix
	\setcounter{table}{0}
	\renewcommand{\thetable}{A\arabic{table}}
	
	\section{Trotterization and gate implementation for the LEO method}
	\label{app:circuit}
	Here we detail the numerical simulation and experimental implementation procedures for realizing LEO-assisted adiabatic speedup on IBM superconducting quantum processors.
	
	\subsection{Initial State Preparation}
	On the IBM quantum processor, each qubit is initialized to the ground state $|0\rangle$ following every reset operation. To meet the initial condition required for adiabatic evolution, which demands that the system starts in the instantaneous ground state of the initial Hamiltonian, a Pauli $X$ gate is executed on the target qubit. This operation flips the state to $|1\rangle$, which then serves as the initial state for the subsequent adiabatic evolution.
	
	\subsection{Time Discretization for Digital Quantum Simulation}
	Traditional studies of quantum many-body dynamics typically rely on continuous-time evolution frameworks. In contrast, digital quantum simulation employs a \textit{discrete-time} paradigm constructed from elementary unitary gates and projective measurements, which is compatible with gate-based superconducting quantum processors.
	
	To realize the time-evolution operator $U(t)$ acting on the initial state $|\psi(0)\rangle$ on IBM quantum hardware, we discretize the total evolution time $t$ into a series of uniform small time steps $\Delta t$. The full evolution operator is then expressed as a product of discrete-time evolution operators:
	\begin{equation}
		U(t) = \underbrace{U(\Delta t) \, U(\Delta t) \cdots U(\Delta t)}_{N\ \text{steps}},
	\end{equation}
	where $N = t/\Delta t$ denotes the total number of discrete time steps.
	
	\subsection{First-Order Suzuki-Trotter Decomposition (Trotterization)}
	The total system Hamiltonian, consisting of the bare adiabatic Hamiltonian $H_0(t)$ and the leakage elimination operator (LEO) control Hamiltonian $H_{\text{LEO}}(t)$, is decomposed into an executable gate sequence using the first-order Suzuki-Trotter approximation. This step is critical for mapping the continuous Hamiltonian dynamics to a discrete gate circuit.
	
	The total evolution operator over the total evolution time $T_{\text{tot}}$ is approximated as:
	\begin{equation}
		U(T_{\text{tot}}) \approx \prod_{k=1}^N e^{-iH_0(t_k)\Delta t} e^{-iH_{\text{LEO}}(t_k)\Delta t},
	\end{equation}
	where $t_k = k\Delta t$ represents the time at the $k$-th step, and the product runs sequentially over all discrete time steps.
	
	\begin{figure}[htbp]
		\centering
		\includegraphics[width=1\columnwidth]{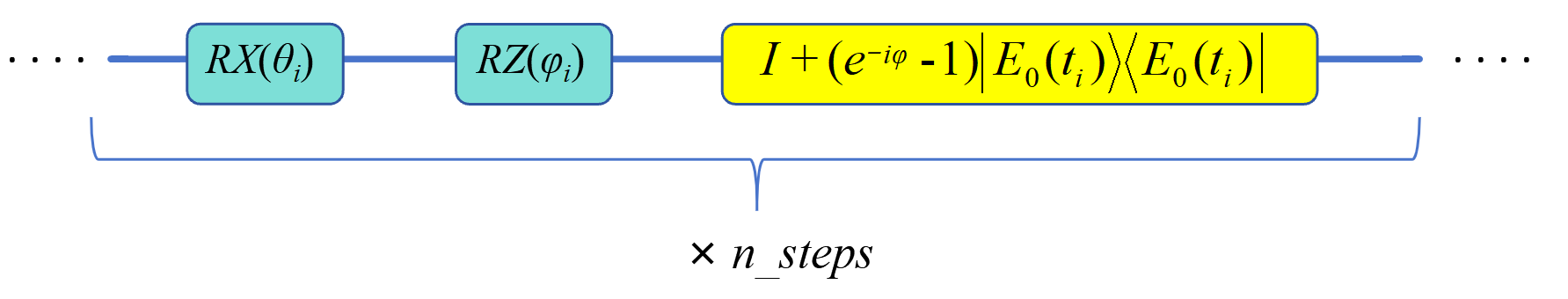}
		\caption{Circuit diagram for a single Trotter step with LEO control.}
		\label{fig:5}
	\end{figure}
	
	\subsection{Decomposition into Native Basis Gates}
	After Trotterization, the resulting unitary operators $e^{-iH_0(t_k)\Delta t}$ and $e^{-iH_{\text{LEO}}(t_k)\Delta t}$ are further decomposed into the native basis gates supported by IBM superconducting quantum processors (including single-qubit gates $I$, $RX$, $RZ$, $SX$, $X$, and two-qubit gates $\text{CZ}$ and $\text{RZZ}$). During compilation, the circuit optimization level is set to $\mathrm{optimization\_level}=0$ to prevent circuit length optimization, thereby making the accumulation of noise more pronounced. This decomposition and compilation strategy ensures that the entire evolution circuit can be directly executed on real quantum hardware without introducing additional approximation errors.
	
	\subsection{Quantum Measurement and Fidelity Analysis}
	Upon completion of the discretized time evolution, we perform projective measurements in the computational basis to evaluate the adiabatic fidelity. The key dynamical quantity, namely the instantaneous ground-state population $F(t)$, is calculated as:
	\begin{equation}
		F(t) = \left| \langle E_0(t) | \psi(t) \rangle \right|^2,
		\label{eq:fidelity}
	\end{equation}
	where $|\psi(t)\rangle$ is the \textit{actual} time-evolved quantum state of the two level system (obtained from simulation or hardware measurement), and $|E_0(t)\rangle$ denotes the \textit{instantaneous ground state} of the total Hamiltonian $H(t) = H_0(t) + H_{\text{LEO}}(t)$ at time $t$. This quantity quantifies the fidelity of adiabatic evolution under LEO control.
	\section{Comprehensive noise model}
	\label{app:noise}
	
	In this work, in order to make the numerical simulation results as close as possible to the behavior of a real quantum computer, the system is treated as an open quantum system interacting with its environment, and quantum noise channels are introduced to characterize the resulting non-unitary evolution. In general, the dynamics of an open quantum system can be described by a completely positive trace-preserving (CPTP) map acting on the density matrix $\rho$, which can be written as
	\begin{eqnarray}
		\mathcal{E}(\rho) = \sum_i K_i \rho K_i^\dagger, \quad \sum_i K_i^\dagger K_i = I,
	\end{eqnarray}
	where $K_i$ are Kraus operators describing the interaction between the system and the environment. In this work, three types of noise are mainly considered: thermal relaxation noise, dephasing noise, and depolarizing noise. The thermal relaxation noise is determined by the relaxation time $T_1$ and the dephasing time $T_2$, which respectively describe the energy decay process from the excited state to the ground state and the loss of phase information; the dephasing noise can be regarded as the phase damping component of the thermal relaxation process, leading to an exponential decay of the off-diagonal elements of the density matrix; the depolarizing noise corresponds to the random application of Pauli operators $X$, $Y$, or $Z$ on the quantum state with a certain probability during the evolution. For the single-qubit case, it can be expressed as
	\begin{eqnarray}
		\mathcal{E}_{\mathrm{depol}}(\rho) = (1 - p)\rho + \frac{p}{3}\left( X\rho X + Y\rho Y + Z\rho Z \right),
	\end{eqnarray}
	where $p$ is the depolarizing probability. This process is equivalent to the system remaining unchanged with probability $1-p$, and undergoing the three types of Pauli errors with equal probability $p$, thereby gradually losing its coherence and purity and eventually approaching a maximally mixed state.
	
	The total noise channel $\mathcal{E}$ is constructed by composing different physical noise processes,
	\begin{eqnarray}
		\mathcal{E} = \mathcal{E}_{\mathrm{thermal}} \circ \mathcal{E}_{\mathrm{depol}},
	\end{eqnarray}
	which is equivalent to introducing environmental noise after each quantum gate operation, thus extending the system dynamics from ideal unitary evolution to non-unitary evolution in the density matrix formalism. In the practical implementation, the noise model (NoiseModel) provided by Qiskit Aer is used to construct these noise channels and automatically insert them after single-qubit gates, in order to simulate the error sources in real quantum hardware.
	
	In particular, the parameters of thermal relaxation noise and dephasing noise are obtained from the real-time calibration data of the \texttt{ibm\_marrakesh} processor, corresponding to the relaxation time $T_1$ and the dephasing time $T_2$ of the qubits, thereby realistically reflecting the energy relaxation and phase damping processes in the hardware. Based on this, when the fidelity obtained from simulations considering only these noises is significantly higher than that measured on the real quantum computer, depolarizing noise is introduced to further refine the model. The strength parameter of the depolarizing noise is determined by fitting, namely by comparing the time-dependent fidelity curves obtained from numerical simulation and those measured on the real quantum computer, and selecting the parameter that minimizes the deviation between the two, thereby achieving an effective characterization of the real noise. The resulting model should  therefore be regarded as a calibrated phenomenological model rather than a complete microscopic reconstruction of all hardware  error channels.
	
	All noise parameters used in the simulations are as follows: In Figure~\ref{fig:1}, $T_1 = 180.86\,\mu\text{s}$, $T_2 = 94.15\,\mu\text{s}$, $p_x = p_y = p_z = 0.0005$, and the numbers of shots are 1000, 1000, and 500 for $T_{\mathrm{tot}} = 1, 5, 15$, respectively. In Figure~\ref{fig:2}, $T_1 = 180.86\,\mu\text{s}$, $T_2 = 94.15\,\mu\text{s}$, with no depolarizing noise, and the number of shots is 1000. In Figure~\ref{fig:3}, the simulations use the noise parameters $T_1 = 196.04\,\mu\text{s}$, $T_2 = 98.82\,\mu\text{s}$, with $1000$ shots and no depolarizing noise. In Figure~\ref{fig:4}, $T_1 = 196.04\,\mu\text{s}$, $T_2 = 98.82\,\mu\text{s}$, with 1000 shots and no depolarizing noise. The single-qubit gate duration is 36 ns throughout.
	
	\section{Pulse Optimization}
	\label{app:optimization}
	
	\subsection{Fourier parameterization of the control phase}
	
	The total evolution time is divided into $M$ optimization segments, each of duration
	\begin{equation}
		T_p = \frac{T_{\mathrm{tot}}}{M},
	\end{equation}
	with $M=10$ in this work. Unlike the original ideal rectangular pulses, whose physical oscillation period is fixed by the parameter $\tau$, the segmentation here is introduced solely for optimization purposes: within each segment, the control field shares one set of Fourier coefficients.
	
	Specifically, the control function $c(t)$ is directly parameterized using a truncated odd-harmonic Fourier sine expansion,
	\begin{equation}
		c^{(p)}(t)
		=
		\sum_{j=0}^{K-1}
		a_j^{(p)}
		\sin\!\big[(2j+1)\Omega_\tau t\big],
	\end{equation}
	where
	\begin{equation}
		\Omega_\tau = \frac{2\pi}{\tau},
	\end{equation}
	$K$ is the number of harmonics, and $a_j^{(p)}$ are the Fourier coefficients to be optimized within the $p$-th optimization segment. Only odd harmonics are included in the expansion to preserve the oscillatory zero-area character of the control pulses while avoiding asymmetric DC offsets. In this work, $K=10$ harmonics are adopted.
	
	After constructing the control field $c(t)$, the discrete accumulated phases used in the Trotterized evolution are obtained through numerical integration,
	\begin{equation}
		\phi_k \approx c_k \Delta t,
	\end{equation}
	where
	\begin{equation}
		\Delta t = \frac{T_{\mathrm{tot}}}{N_{\mathrm{steps}}},
	\end{equation}
	is the Trotter time step.
	
	\subsection{Greedy sequential period-wise optimization}
	
	A greedy sequential segment-wise optimization method is employed. The optimization proceeds sequentially from the first segment, where the Fourier coefficients of previously optimized segments are kept fixed during subsequent optimizations. When optimizing the $p$-th segment, the coefficients of segments $0$ through $p-1$ have already been determined and remain fixed, while the coefficients of segments $p+1$ through $M-1$ are temporarily set to zero. Only the $K$ Fourier coefficients of segment $p$ are treated as free optimization parameters. The objective function for each subproblem is defined as the adiabatic fidelity at the end of the $p$-th segment, evaluated through noisy quantum simulation.
	
	\subsection{Bayesian optimization}
	
	For each subproblem, Bayesian optimization is employed to search for the optimal Fourier coefficients $a_j^{(p)}$. The optimization is initialized using Fourier coefficients with amplitudes determined by the ideal square-wave pulse, and each coefficient is allowed to vary within a bounded local neighborhood around its corresponding initial value. Bayesian optimization employs a Gaussian-process surrogate model of the objective function and can efficiently identify near-optimal solutions with a limited number of noisy function evaluations. Once the optimal coefficients for a given optimization segment are obtained, they are accepted and fixed, and the algorithm proceeds sequentially to optimize the next segment. This strategy offers several advantages: the dimensionality of each subproblem remains low, allowing Bayesian optimization to operate efficiently; the greedy sequential strategy provides an effective optimization heuristic, with stable improvement observed in our numerical simulations; and the method naturally accommodates time-dependent noisy dynamics, since the optimization of later segments is conditioned on the actual evolution path generated by the previously optimized segments.
	
	\subsection{Implementation parameters}
	
	In the numerical experiments presented in this work, the total evolution time is $T_{\mathrm{tot}} = \pi$, and the number of Trotter steps is $n_{\mathrm{steps}} = 160$. The total evolution is divided into $M = 10$ optimization segments, with each segment containing $16$ Trotter steps. The control field in each segment is expanded using $K = 10$ odd-harmonic Fourier sine components, resulting in a total of $100$ optimization parameters. Bayesian optimization is performed sequentially segment by segment using a Gaussian-process surrogate model with a Matérn-$5/2$ kernel. The Gaussian process employs normalized target values and includes a small observation noise term for numerical stability. For each optimization segment, $30$ Bayesian optimization evaluations are performed, including $10$ initial evaluations initialized from the Fourier coefficients of the ideal square-wave pulse. The acquisition function is chosen as expected improvement (EI) with exploration parameter $\xi = 0.5$. The acquisition-function optimization is carried out using a sampling-based strategy with $10^4$ candidate points. During the optimization, the Fourier coefficients are restricted to vary within a bounded local neighborhood around the corresponding ideal coefficients, enabling noise-adapted waveform optimization while preserving the overall structure of the ideal zero-area pulse.

	\bibliography{refs.bib}
\end{document}